\documentclass{elsart}
\usepackage{graphicx}
\begin{document}
\begin{frontmatter}
\title{New light on electromagnetic corrections to the scattering parameters
obtained from experiments on pionium }
\author[ZH]{A. Gashi}
\author[AA]{G.C. Oades\thanksref{cor}}
\author[ZH]{G. Rasche}
\author[CA]{W.S. Woolcock}

\address[ZH]{Institut f\"{u}r Theoretische Physik der Universit\"{a}t,
Winterthurerstrasse 190, CH-8057 Z\"{u}rich, Switzerland}
\address[AA]{Institute of Physics and Astronomy, Aarhus University,
DK-8000 Aarhus C, Denmark}
\address[CA]{Department of Theoretical Physics, IAS,
The Australian National University, Canberra, ACT 0200, Australia}
\thanks[cor]{Corresponding author. Electronic mail: gco@ifa.au.dk;
Tel: +45 8942 3653;  Fax: +45 8612 0740}

\begin{abstract}

We calculate the electromagnetic corrections needed to obtain isospin
invariant hadronic pion-pion $s$-wave scattering lengths $a^0$, $a^2$
from the elements $ a_{cc}, a_{0c}$ of the $s$-wave scattering matrix
for the $(\pi^+\pi^-, \pi^0\pi^0)$ system at the $\pi^+\pi^-$ threshold.
The latter can be extracted from experiments on the $\pi^+\pi^-$ atom
(pionium). Our calculation uses energy independent hadronic pion-pion
potentials $V^0, V^2$ that satisfactorily reproduce the low-energy phase
shifts given by two-loop chiral perturbation theory, with the hadronic
mass of the pion taken first as the charged pion mass and then as the
neutral pion mass. We also take into account an  important relativistic
effect whose inclusion influences the corrections considerably.
\end{abstract}


\end{frontmatter}
\section{Introduction}

The extraction of the $s$-wave pion-pion  scattering lengths $a^I ,I=0,2$, 
from the results of the DIRAC experiment  currently in progress at CERN
\cite{1} requires a knowledge of the electromagnetic corrections
$a_{0c}-\frac{\sqrt{2}}{3}(a^2-a^0)$ and $a_{cc}-(\frac{2}{3}a^0+
\frac{1}{3}a^2)$, where $a_{cc}$ and $a_{0c}$ are elements of the well
known matrix $\bf{K}$ of scattering theory in the $s$-wave  for the
two-channel $(\pi^+\pi^-, \pi^0\pi^0)$ system at the $\pi^+\pi^-$ threshold.
(The subscript $c$ refers to the $\pi^+\pi^-$ channel, $0$ to the 
$\pi^0\pi^0$ channel.) 
Previous attempts to calculate these corrections using a potential model
have been reported in Refs.\cite{2,3}. 
In this paper we improve on the calculation of Ref.\cite{3} by using new
hadronic potentials obtained by fitting the phase shifts given by two-loop
chiral perturbation theory (ChPT) and by including a relativistic
modification which turns out to have a significant effect even in the
threshold situation we are considering.

This relativistic modification emerged as an important issue in the
calculation of electromagnetic corrections for the analysis of low energy
$\pi^-p$ scattering data \cite{4}. 
The treatment of the two-channel ($\pi^-p, \pi^0n$) system closely
parallels that of the $(\pi^+\pi^-, \pi^0\pi^0)$ system and the coupled
relativised Schr\"{o}dinger equations (RSEs) for the two systems are
formally identical. The two-channel RSEs that we now use to model the
$s$-wave of the physical $(\pi^+\pi^-, \pi^0\pi^0)$ system are therefore
given by  Eq.(21) of Ref.\cite{4}:
\begin{equation}
\left \{\,\mathbf{1}_2\frac{d^2}{dr^2}+\mathbf{Q}^2-2\mathbf{m}\mathbf{f}\mathbf{V}(r) \right \} \mathbf{u}(r)=\mathbf{0},
\label{eq:1}
\end{equation}
where
\begin{equation}
\mathbf{m}= \left( \begin{array}{cc}
\frac{1}{2}\mu_c & 0  \\
0 & \frac{1}{2}\mu_0 \end{array} \right) ,
\mathbf{Q}= \left( \begin{array}{cc}
q_c & 0  \\
0 & q_0 \end{array} \right) , 
\mathbf{f}= \left( \begin{array}{cc}
f_c & 0  \\
0 & f_0 \end{array} \right)  .
\label{eq:2}
\end{equation}
In Eq.(\ref{eq:2}), $\mu_c$ and $\mu_0$ are the physical 
masses of $\pi^{\pm}$ and  $\pi^0$ respectively, while $q_c$ and
$q_0$ are the c.m. momenta of the two channels. Further,
\begin{equation}
f_c=\frac{W^{2}-2\mu_c^2}{\mu_cW} \, , \, f_0=\frac{W^{2}-2\mu_0^2}{\mu_0W} ,
\label{eq:3}
\end{equation}
where $W$ is the total energy in the c.m. frame. Note that $f_c$ and
$f_0$ are $1$ at the respective thresholds $W=2\mu_c$ and  $W=2\mu_0$
and increase as $W$ increases.

Eq.(\ref{eq:1}) now replaces Eq.(8) of Ref.\cite{3}. The inclusion of the
factor $\bf {f}$ is of crucial importance for the calculation of the
electromagnetic corrections for both the $(\pi^+\pi^-, \pi^0\pi^0)$
and $(\pi^-p, \pi^0n)$ systems. The first part of Section 3 of Ref.\cite{4}
gives the arguments for its inclusion in the RSEs for the latter system.
The potential matrix $\bf{V}$ has the form 
\begin{equation}
\mathbf{V}={\mathbf{V}}^{em}+\mathbf{V}^h .
\label{eq:4}
\end{equation}
The electromagnetic potential matrix has only one nonzero entry
${(\bf{V}}^{em})_{cc}=V^{em}$. In the 
notation of Ref.\cite{4}, $V^{em}$ contains only $V^{pc}$ (point charge
Coulomb) and $V^{ext}$ (which takes account of the extended charge
distributions); the effect of $V^{rel}$ and $V^{vp}$ 
(vacuum polarisation) on the positions and widths of the levels of pionium
is treated separately. 

The hadronic potential matrix $\bf {V^{\it h}}$ is assumed to be isospin
invariant, in accordance with the arguments of Gasser and Leutwyler
\cite{5,6}, which show that there is practically no isospin 
breaking of the purely hadronic pion-pion scattering amplitudes. Thus 
\begin{equation}
\mathbf{V}^h=\left( \begin{array}{cc}
\frac{2}{3}V^0+\frac{1}{3}V^2 & \frac{\sqrt{2}}{3}(V^2-V^0) \\
\frac{\sqrt{2}}{3}(V^2-V^0) & \frac{1}{3}V^0+\frac{2}{3}V^2 \end{array} \right) .
\label{eq:5}
\end{equation}
The determination of the hadronic potentials will be considered in
Section 2. Once they are chosen, the coupled 
equations (\ref{eq:1}) are integrated numerically to give $\bf{ K}$
as a function of $W$. Extrapolation to $W=2\mu_c$ gives the matrix 
\begin{equation}
\mathbf{a}=\mathbf{K} (2\mu_c) ,
\label{eq:6}
\end{equation}
whose elements $a_{cc}$, $a_{0c}$ can be obtained from experiments on
pionium. The determination of these quantities from pionium  data is
reviewed at the beginning of Section 3, which then sets out the results
of our calculation of the electromagnetic corrections needed to obtain
the hadronic pion-pion scattering lengths $a^0$,$a^2$ from $a_{cc}$ and
$a_{0c}$. In Section 4 we compare our results with those of ChPT.

\section{The hadronic pion-pion potentials}

We now need to model the hadronic situation and determine the potentials
$V^0$, $V^2$. Chiral perturbation theory (ChPT) has proved a fruitful
effective theory of low energy pion-pion scattering that incorporates the
essential constraint of chiral invariance. From the point of view of ChPT,
the hadronic mass $\mu$ of the pion (assumed the same for 
$\pi^{\pm}$ and $\pi^0$) is an adjustable parameter and ChPT is able to
generate hadronic phase shifts $\delta^0$, $\delta^2$ that depend on the
value of $\mu$ that is chosen. This procedure is complicated by the need
to fix certain low energy constants using experimental data obtained with
physical pions. Nevertheless, Colangelo \cite{7} has  provided us with
hadronic $s$-wave phase shifts $\delta^I(\mu;W)$, $I=0$,$2$, obtained 
from two-loop ChPT as functions of $W$ for a region above the threshold
$W=2\mu$, for $\mu=\mu_c$ and for $\mu=\mu_0$. 

What we have been able to do is to find potentials $V^0$, $V^2$ that are
independent of both the energy $W$ and the 
hadronic mass $\mu$, which reproduce quite well the two-loop ChPT phase
shifts of Ref.\cite{7} for $\mu=\mu_c$ and $\mu=\mu_0$ via the RSEs
\begin{equation}
\left (\,\frac{d^2}{dr^2}+q_c^2-\mu_cf_{c}V^I(r)\,\right ) \,u(r)=0 ,
\label{eq:7}
\end{equation}
\begin{equation}
\left (\,\frac{d^2}{dr^2}+q_0^2-\mu_0f_{0}V^I(r)\,\right ) \,u(r)=0 ,
\label{eq:8}
\end{equation}
with $I=0$, $2$. The important difference from Eq.(6) of Ref.\cite{3}
(which applies to $\mu=\mu_c$ only) is the presence 
of the factors $f_c$ and $f_0$ in Eqs.(\ref{eq:7}) and (\ref{eq:8}).
Compared with the work of Ref.\cite{3}, we were also able to use our 
experience of constructing potentials for calculating electromagnetic
corrections in low energy pion-nucleon scattering to 
obtain physically more realistic potentials than the double square
wells used earlier.

\begin{figure}
\begin{center}
\includegraphics[height=0.65\textheight,angle=0]{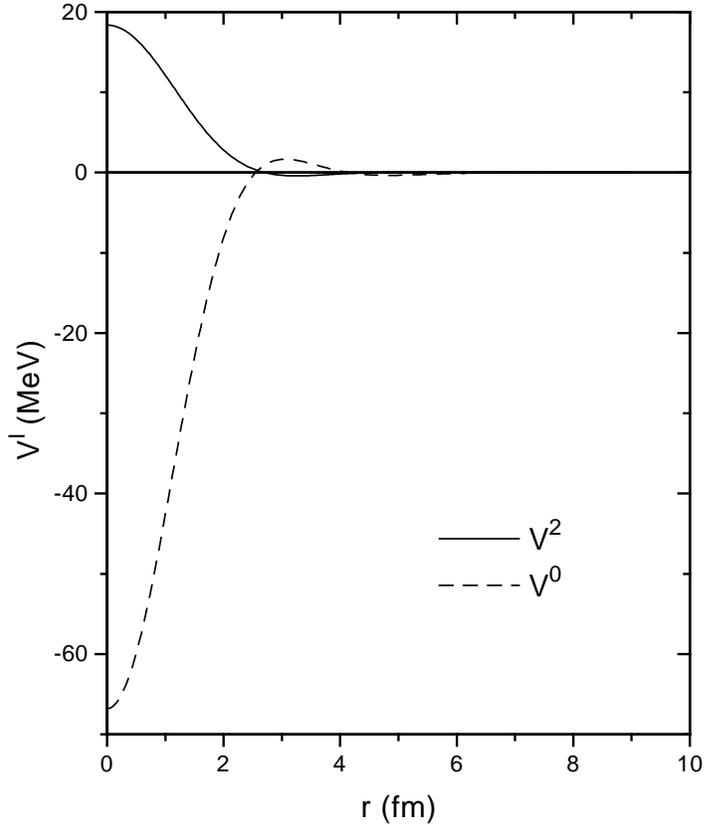}
\caption{The potentials $V^0(r)$ and $V^2(r)$ that give the best fit
to the two-loop ChPT phase shifts of Ref.\cite{7} for hadronic masses
$\mu_c$ and $\mu_0$.}
\label{fig:1}
\end{center}
\end{figure}

\begin{figure}
\begin{center}
\includegraphics[height=0.65\textheight,angle=0]{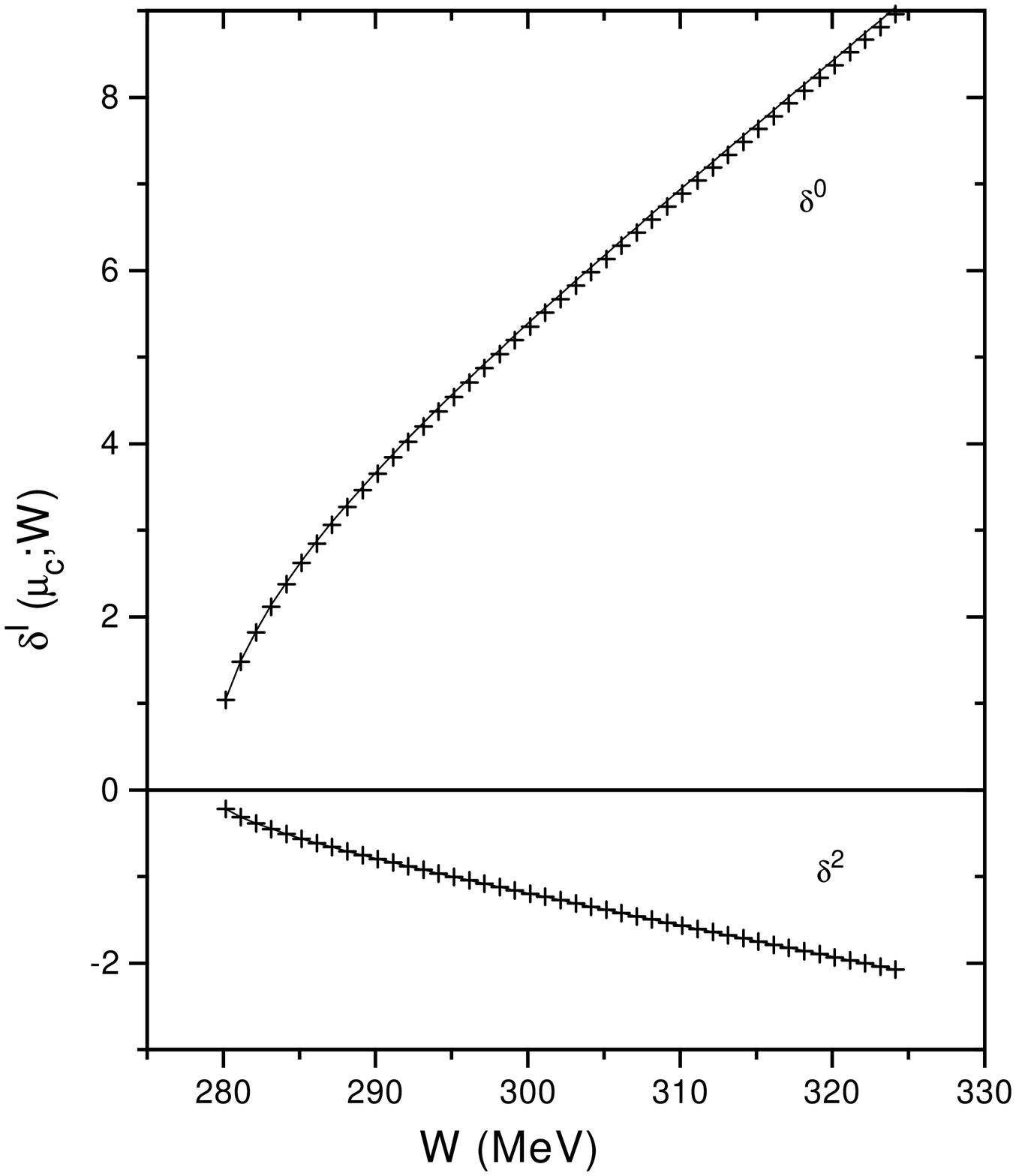}
\caption{The phase shifts $\delta^0(\mu_c;W)$ and $\delta^2(\mu_c;W)$
(in degrees) for hadronic mass $\mu_c$ given by two-loop ChPT (crosses)
and by the potentials $V^0$ and $V^2$ of Fig.\ref{fig:1} (solid curves).}
\label{fig:2}
\end{center}
\end{figure}

\begin{figure}
\begin{center}
\includegraphics[height=0.65\textheight,angle=0] {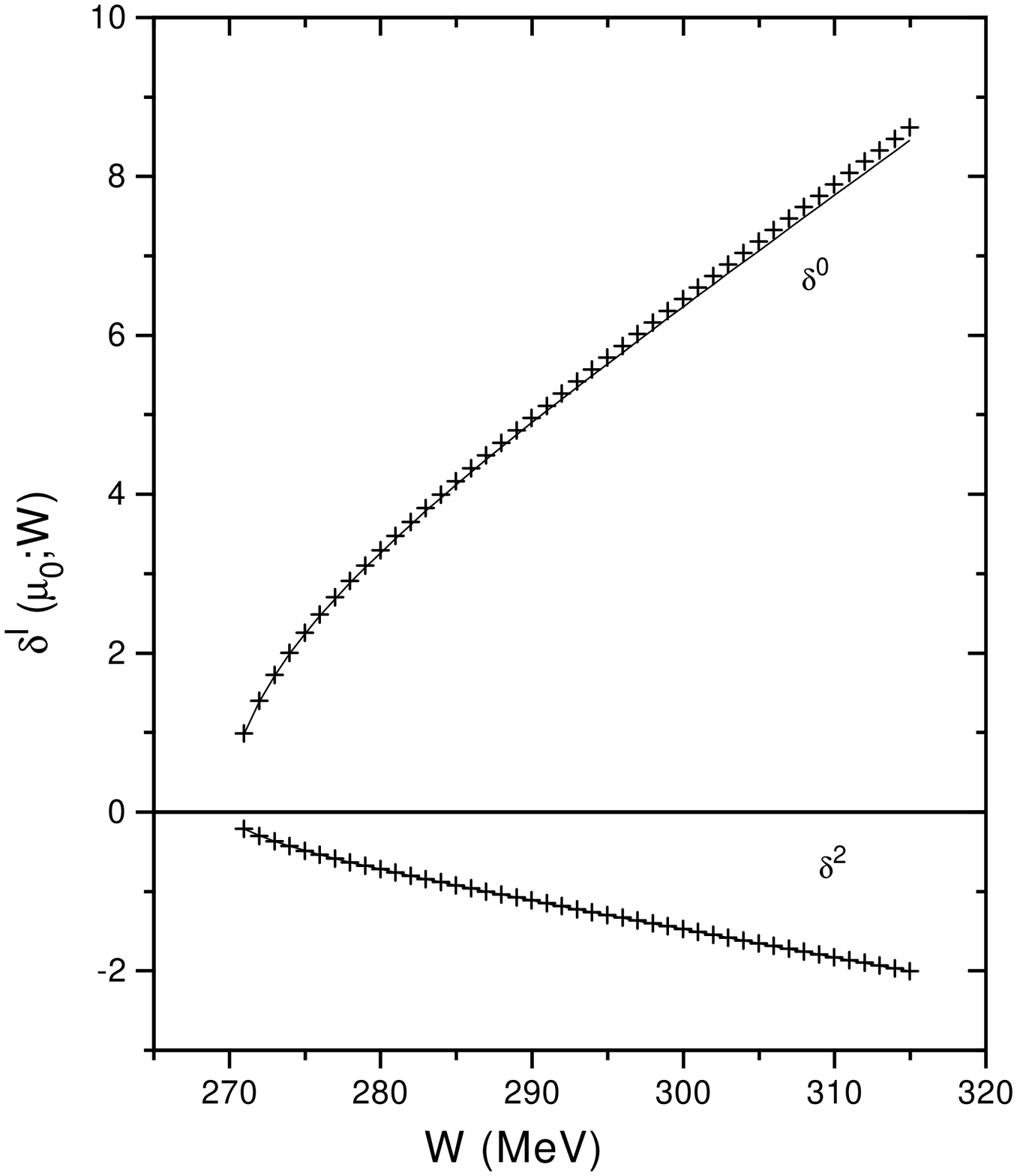}
\caption{The phase shifts $\delta^0(\mu_0;W)$ and $\delta^2(\mu_0;W)$
(in degrees) for hadronic mass $\mu_0$ given by two-loop ChPT (crosses)
and by the potentials $V^0$ and $V^2$ of Fig.\ref{fig:1} (solid curves).}
\label{fig:3}
\end{center}
\end{figure}

We found that the best fit to the phase shifts was obtained when the
range parameter in these potentials was 1.5 fm. Each potential contains
three further parameters that were varied to obtain the best fit. Full
details of the parameterisation are given in Ref.\cite{8}. We found that
it is possible to construct for $I=0$, $2$ potentials $V^I(r)$ that give
a quite good fit to the two-loop ChPT phase shifts 
$\delta^I(\mu;W)$ for $\mu=\mu_c$ and $\mu=\mu_0$, using Eqs.(\ref{eq:7})
and (\ref{eq:8}) to generate phase shifts by integrating the 
regular solution outwards from the origin. These potentials are plotted
in Fig.\ref{fig:1}. It is impossible to obtain 
a satisfactory fit to the phase shifts for both values of $\mu$ when the
factors $f_c$ and $f_0$ are omitted from Eqs.(\ref{eq:7}) and 
(\ref{eq:8}); this is further significant evidence for the need to
include them. What we have been able to do is to capture some of the 
dynamical behaviour of ChPT by means of a simple potential model involving
potentials independent of the energy and the 
hadronic mass. In Figs.\ref{fig:2} and \ref{fig:3} we show how well
the potentials reproduce the ChPT phase shifts for $\mu=\mu_c$ and 
$\mu=\mu_0$ respectively. Table \ref{tab:1} gives a comparison of the
hadronic scattering lengths given by two-loop ChPT and by our potentials.

\begin{table}
\begin{center}
\caption{The hadronic pion-pion scattering lengths for total isospin
$I=0,2$ and hadronic mass $\mu=\mu_c, \, \mu_0$ 
given by two-loop ChPT (Ref.\cite{7}) and by the potentials $V^0$ and
$V^2$ of Fig.\ref{fig:1} (all results in fm).}
\label{tab:1}
\begin{tabular}{|c|c|c|}
\hline
  & two-loop ChPT & our potentials \\\hline
$a^0(\mu_c)$ & 0.3055 & 0.3004 \\
$a^0(\mu_0)$ & 0.2930 & 0.2892 \\
$a^2(\mu_c)$ & -0.0632 & -0.0632 \\
$a^2(\mu_0)$ & -0.0614 & -0.0613 \\ \hline
\end{tabular}
\end{center}
\end{table}

\section{From pionium data to hadronic pion-pion scattering lengths}

The first step is the extraction of the quantities $a_{cc}$, $a_{0c}$
from the experimental values of the lifetime  $\tau$ of the $1s$ level
of pionium and of the difference $\Delta W^{had}$ between the energies
of the $2s$ and $2p$ levels. This is considered in Section 2 of
Ref.\cite{3}, with the results
\begin{equation}
a_{0c} (\rm fm)=-0.3794\, \,[\Gamma (\rm eV)]^{1/2} ,
\label{eq:9}
\end{equation}
\begin{equation}
a_{cc} (\rm fm)=-0.4167\,\,\Delta W^{had}(\rm eV) .
\label{eq:10}
\end{equation}
In terms of the lifetime $\tau$, Eq.(\ref{eq:9}) becomes    
\begin{equation}
a_{0c} (\rm fm)=-0.3078\,\, [\tau (\rm fs)]^{-1/2} .
\label{eq:11}
\end{equation}
The slight differences between the numerical constants of
Eqs.(\ref{eq:9}) and (\ref{eq:10}) and those in Ref.\cite{3} come
from a new value of $a_{cc}$ and the direct inclusion of the effect
of vacuum polarisation in Eq.(\ref{eq:9}). These results are based
on the work of Ref.\cite{9}, which uses general principles of
scattering theory, in particular the analytic continuation of $\bf {K}$
below the $\pi^+\pi^-$ threshold. The basic formalism goes back to
Hamilton, {\O}verb\"{o} and Tromborg \cite{10} and was worked out
systematically for multichannel situations by Rasche and Woolcock
\cite{11}. It is independent of the potential model, which we  use
only to calculate electromagnetic corrections.

It is important to observe that the expressions for the hadronic shifts
and widths of the levels of hadronic atoms obtained in Ref.\cite{9} are
given in terms of the elements of $\bf{a}$ as defined in Eq.(\ref{eq:6}).
As shown in Ref.\cite{9} there is no significant error involved in
evaluating $\bf{K}$ at the threshold of the charged channel instead of
at the pole position corresponding to the bound state. Thus the
extraction of $a_{cc}$ and $a_{0c}$ from experimental data is
completely separated from the question of the electromagnetic
corrections that relate these two numbers to the hadronic $s$-wave
scattering lengths $a^I,\,\,I=0,2$. This separation is valuable from
a conceptual point of view.

In order to compare our results with field theoretical treatments of
the width $\Gamma$, it is necessary to give our definitions of
two-channel scattering amplitudes and to establish the connection
with the standard amplitudes used in work on ChPT. We confine ourselves
again to the $s$-wave.  Starting from the unitary matrix $\bf{S}$,
we define matrices $\bf{T}$, $\bf{K}$ and $\bf{A}$ as follows:
\begin{equation}
\mathbf {T}= - i(\mathbf {S}- \mathbf {1}_2)(\mathbf {S}+ \mathbf {1}_2)^{ - 1} ,
\label{eq:12}
\end{equation}
\begin{equation}
\mathbf{K}^{-1}=\left( \begin{array}{cc}
C_0(\eta _-) & 0 \\
0 & 1 \end{array} \right)\mathbf{Q}^{1/2}\mathbf{T}^{-1}\mathbf{Q}^{1/2}\left( \begin{array}{cc}
C_0(\eta _-) & 0 \\
0 & 1 \end{array} \right)+\left( \begin{array}{cc}
-\beta h(\eta_-) & 0 \\
0 & 0 \end{array} \right) ,
\label{eq:13}
\end{equation}
\begin{equation}
\mathbf {A}^{-1}= \mathbf{K}^{-1} -i\mathbf {Q}^{-1} .
\label{eq:14}
\end{equation}
The matrix $\bf{Q}$ is given in Eq.(\ref{eq:2}), while $\beta=\alpha \mu_c$, $\eta_-=-\beta /2q_c$,

\[C_0^2(\eta)=\frac{2\pi \eta}{exp(2\pi \eta)-1} \, , \, h(\eta)=-ln\mid\eta\mid +\Re\psi (1+i\eta) \, .
\]
 Eq.(\ref{eq:13}) is the explicit definition of $\bf{K}$, in which the
specific low energy behaviour of $\bf{T}$ induced by the Coulomb
interaction is taken into account. The complex quantity $A_{cc}$ plays
the key role in the formalism of Ref.\cite{9}.

If we denote by $\bf{A}^\chi$ the $s$-wave part of the matrix of
scattering amplitudes used in work on ChPT, the connection with the
matrix $\bf{A}$ of Eq.(\ref{eq:14}) is
\begin{equation}
\mathbf{A}={(32\pi W)}^{-1}
\left( \begin{array}{cc}
\sqrt{2} & 0 \\
0 & 1 
\end{array} \right) \bf{A}^\chi
\left( \begin{array}{cc}
\sqrt{2} & 0 \\
0 & 1 
\end{array} \right) .
\label{eq:15}
\end{equation}
Note that $\bf{A}$ has the dimension of a length, while  $\bf{A}^\chi$
is dimensionless. The most satisfactory field theoretical treatment
of pionium is given in Gall $et\, al$.\cite{12}, which uses QCD
(including photons) and works with the low energy expansion provided
by ChPT. Ref.\cite{12} also gives references to and discusses other
field theoretic methods of treating hadronic atoms. In Ref.\cite{13},
Gasser, Lyubovitskij and Rusetsky apply the formalism of Ref.\cite{12}
to the numerical calculation of $\Gamma$. A more detailed account of
their work is given in Ref.\cite{14}. The expression for $\Gamma$
given in Refs.\cite{12,13,14} involves the real part of an amplitude
$A_{thr}^{+-00}$, from which the specific low energy behaviour due to
the Coulomb interaction has been removed. Using Eq.(\ref{eq:15}), we have
\begin{equation}
A_{0c}(2\mu_c)= \sqrt{2}{(64\pi \mu_c)}^{-1}A_{thr}^{+-00}..
\label{eq:16}
\end{equation}

If we compare the expression for $\Gamma$ given in Refs.\cite{12,13,14}
with our result in Eq.(9) of Ref.\cite{3}, on which Eq.(\ref{eq:9})
above is based, we find that a term proportional to ln$\,{\alpha}$ and
the purely hadronic scattering lengths appear in the quantity $K$ given
in Eq.(4) of Ref.\cite{12}, while our result has no term involving
ln$\,{\alpha}$ and contains the scattering parameters that include the
effect of electromagnetic interaction. As Ivanov $et\,al.$ [15] have
pointed out, these differences come from the same source: the
treatment of Coulomb photons. In our method the Coulomb potential is
included in the RSEs (\ref{eq:1}), which are solved exactly. The
nonanalytic behaviour in $\alpha$ is implicitly included in our
electromagnetic corrections. The treatments based on scattering
theory \cite{9} on the one hand and on field theory \cite{12,13,14}
on the other therefore lead in the end numerically to the same results.

To understand the electromagnetic corrections we must distinguish
between the physical situation, in which the electromagnetic interaction
is present and the pions have their observed masses, and the hadronic
situation, in which the electromagnetic interaction is switched off.
The pioneering work of Gasser and Leutwyler \cite{5,6}, already
used in Section 1, shows that, at first order in the low energy
expansion of ChPT, in the hadronic situation the only isospin
breaking effect due to the difference between the quark masses
$m_u$ and $m_d$ is a very small shift in the $\pi^0$ mass. It is
therefore a very good approximation to take the hadronic situation
as one in which all the pions have the same mass $\mu$ and the
$s$-wave amplitudes are isospin invariant.

As noted in Section 2, for ChPT at the hadronic level, the value of
$\mu$ is a parameter that is not determined by the theory.
(This does not mean that its value can be chosen completely
arbitrarily.) Thus ChPT contains dynamical information about
$s$-wave hadronic $\pi\pi$ scattering for a whole range of hadronic
masses of the pions. The potentials constructed in Section 2 capture
a lot of this dynamical information for a range of masses around
$\mu_0$. They were constructed in order to be able to incorporate
the effects of the electromagnetic interaction (the Coulomb potential
for extended charges and the difference between  $\mu_0$ and $\mu_c$)
into a potential model of the physical situation that enables the
electromagnetic corrections to be calculated. That model is just the
two-channel RSEs(1); it is simple and hopefully will give a reasonably
reliable estimate of the corrections. 

We now define the hadronic quantities $a_{cc}^h(\mu)$,  $a_{0c}^h(\mu)$ by
\begin{equation}
a_{cc}^h(\mu)=\frac{2}{3}a^0(\mu)+\frac{1}{3}a^2(\mu) \,\, , \, a_{0c}^h(\mu)=\frac{\sqrt{2}}{3}(a^2(\mu)-a^0(\mu)) .
\label{eq:17}
\end{equation} 
As stated in Section 3 of Ref.\cite{3}, numerical experience shows that
the quantities least sensitive to small variations of the hadronic
potentials $V^I$ are
\begin{equation}
\Delta a_{cc}(\mu)=a_{cc}^h(\mu)-a_{cc} \, , \,\Delta_{0c}(\mu)=a_{0c}^h(\mu)/a_{0c}-1  .
¨\label{eq:18}
\end{equation}
The term `electromagnetic corrections' will in future mean
$\Delta a_{cc}(\mu)$ and $\Delta_{0c}(\mu)$. Note that the hadronic
quantities $a^I(\mu), I=0,2\, ,\, a_{cc}^h(\mu)$ and  $a_{0c}^h(\mu)$
depend explicitly on $\mu$. They are given by an equation like
Eq.(\ref{eq:7}) or Eq.(\ref{eq:8}), with $\mu$ replacing $\mu_c$ or
$\mu_0$ and a momentum variable $q(\mu;W)$ corresponding to the
mass $\mu$. However, the physical scattering parameters $a_{cc}$
and $a_{0c}$ are given by the RSEs(\ref{eq:1}) and do not depend
on $\mu$. The $a^I(\mu)$ given by the potentials of Section 2 are
listed in the second column of Table \ref{tab:1} for $\mu=\mu_c$ and
$\mu=\mu_0$. They lead to the results

\begin{equation}
a_{cc}^h(\mu_c)=0.1791\, {\rm fm} \, \, , \,\, a^{h}_{0c}(\mu_c)=-0.1714\, \rm{fm} ,
\label{eq:19}
\end{equation}
\begin{equation}
a_{cc}^h(\mu_0)=0.1721\, {\rm fm} \, \, , \,\, a^{h}_{0c}(\mu_0)=-0.1651\, \rm{fm} .
\label{eq:20}
\end{equation}

The procedure for determining the elements of $\bf{K}$ at the threshold
of the charged channel was described briefly at the end of Section 1.
The coupled RSEs(\ref{eq:1}) were integrated numerically to obtain
$\bf{K}$ as a function of $W$, as described in Section 3 of Ref.\cite{16}.
The matrix $\bf{K}$ was then  extrapolated to $W=2\mu_c$ to obtain
$\bf{a}$. The results are:
\begin{equation}
a_{cc}=0.1773 {\rm fm}  ,  a_{0c}=-0.1726 {\rm fm}  ,  a_{00}=0.0576 {\rm fm} .
\label{eq:21}
\end{equation}
From Eqs.(\ref{eq:19})-(\ref{eq:21}) the electromagnetic corrections
of Eq.(\ref{eq:18}) are then
\begin{equation}
\Delta a_{cc}(\mu_c)=0.0018 {\rm fm} ,  \Delta_{0c}(\mu_c)=-0.0068 ,
\label{eq:22}
\end{equation}
\begin{equation}
\Delta a_{cc}(\mu_0)=-0.0052 {\rm fm} ,  \Delta_{0c}(\mu_0)=-0.0436 .
\label{eq:23}
\end{equation}
Eqs.(\ref{eq:22}) and (\ref{eq:23}) are the final results of our calculations
of the electromagnetic corrections in the potential model. They
supersede the results in Refs.\cite{2,3}, where much more primitive
hadronic potentials were used and the relativistic factors $f_c,f_0$
were not taken into account.

We have given the corrections for two possible values of the hadronic
mass of the pion. However it was also shown by Gasser and Leutwyler
\cite{5,6} that the mass of the pion in the hadronic situation is very
close to $\mu_0$. It follows that the hadronic starting point, with
respect to which the electromagnetic corrections need to be calculated,
has a pion mass $\mu_0$. The results in Eq.(\ref{eq:23}) are therefore
the true electromagnetic corrections. These need always to be calculated
with $\mu_0$ as the hadronic mass of the pion.

\section{Discussion}

We now compare our results with those obtained from the work of Knecht
and Urech \cite{17}, who use the low energy expansion of ChPT and
include the effect of the electromagnetic interaction. At lowest order
in ChPT the hadronic scattering lengths are
\begin{equation}
a^0(\mu_0)=\frac{7\mu_0}{32 \pi F^2}\, \, , \,\, a^2(\mu_0)=-\frac{\mu_0}{16 \pi F^2}\, , 
\end{equation}
giving
\begin{equation}
a^h_{cc}(\mu_0)=\frac{\mu_0}{8 \pi F^2}\, \, , \,\, a^h_{0c}(\mu_0)=-\frac{3 \sqrt{2} \mu_0}{32 \pi F^2}\, \, ,
\end{equation}
where $F$ is the pion decay constant in the chiral limit ($\approx 88$ MeV).
Using Eq.(\ref{eq:15}) above and Eqs.(2.14)-(2.16) of Ref.\cite{17}, we
find that

\begin{equation}
a_{cc}=\frac{(2\mu_c^2-\mu_0^2)}{8 \pi F^2}\mu_c^{-1}\, \, , \,\, a_{0c}=-\frac{ \sqrt{2}(4\mu_c^2- \mu_0^2)}{32 \pi F^2}\mu_c^{-1}\, \, 
\end{equation}
for the physical situation, with the pions having their observed masses.
Combining these results, the lowest order calculation of Ref.\cite{17}
gives the corrections
\begin{equation}
\Delta a_{cc}(\mu_0)=-0.0139 {\rm fm} ,  \Delta_{0c}(\mu_0)=-0.0533 .
\label{eq:24}
\end{equation}
At the one-loop level, only the amplitude for $\pi^+ \pi^- \rightarrow
\pi^0 \pi^0$ is calculated in Ref.\cite{17}. Using Eq.(\ref{eq:15})
again, the results in Eqs.(5.2) and (5.6) of Ref.\cite{17} yield the
quantities
\begin{equation}
 a_{0c}^h(\mu_0)=-0.1570\, {\rm fm} \, \, , \,\, a_{0c}=-0.1675\, {\rm fm}\, ,
\end{equation}
from which we deduce that
\begin{equation}
\Delta_{0c}(\mu_0)=-0.0625 ,
\label{eq:25}
\end{equation}
to be compared with the leading order result in Eq.(\ref{eq:24}).
These numbers have been calculated using the values of the low energy
constants $\bar{l_i}, i=1-4$, and $\mathcal{K}_1^{\pm 0}$, 
$\mathcal{K}_2^{\pm 0}$ given in Ref.\cite{17}.

To compare the results in Eqs.(\ref{eq:23}) (present work) and
(\ref{eq:25}) (calculated from Ref.\cite{17}) we separate the effect
of the pion mass difference from that of the Coulomb interaction.
The correction due to the mass difference only is 
\begin{equation}
\Delta_{0c}^{md}(\mu_0)=-0.0443\, ({\rm present\,\, work}),\,\, \Delta_{0c}^{md}(\mu_0)=-0.0561\, ({\rm Ref.[17]})\, .
\end{equation}
The uncertainty in the second number, based on the quoted uncertainties
in $\bar{l_i}, i=1-4$, is 0.0015. Moreover, the change from the lowest
order result to the one-loop result is small. The number from our
present work is subject to an uncertainty that comes from the possible
difference between the purely hadronic potential that we have used and
the effective hadronic potential in the presence of the electromagnetic
interaction. This difference can be thought of in a field theory picture
as the result of electromagnetic mass differences in the internal lines
of Feynman diagrams. Assuming that these effects are correctly accounted
for in Ref.\cite{17}, it is likely that the ChPT result gives a more
reliable estimate of the mass difference correction than does our
potential model.

On the other hand, due to the large uncertainties in $\mathcal{K}_1^{\pm 0}$,
$\mathcal{K}_2^{\pm 0}$, the Coulomb correction in the ChPT calculation
has a large uncertainty; it is $-0.0064(82)$. The potential model gives
a small correction, +0.0007, with no significant uncertainty. This result
lies within the ChPT error band and removes the uncertainty.

Combining these two results we think that a realistic value of the
total electromagnetic correction is
\begin{equation}
\Delta_{0c}(\mu_0)=-0.055(5) .
\label{eq:26}
\end{equation}
The error given in Eq.(\ref{eq:26}) is a generous one; it may be smaller.
There is some uncertainty due to the neglect of higher orders in the
chiral expansion and we have allowed the possibility of a larger
uncertainty from the constants $\bar{l_i},\, i$=1-4. Since the DIRAC
experiment \cite{1} aims to deliver a result for $\tau$ that will give
a value of $a_{0c}$ with an error of 5 $\%$, the uncertainty in
$\Delta_{0c}({\mu}_0)$ is much smaller than the expected experimental error.

This means that, if the DIRAC experiment achieves its proposed accuracy,
it will be able to give a result for $a_{0c}^{h}(\mu_0)$ with an accuracy
comparable with that of the present estimates from ChPT. 
Recent work of Colangelo, Gasser and Leutwyler \cite{18}, in which
two-loop ChPT is combined with the Roy equations and phenomenological
data from pion-pion scattering, gives $a^0-a^2$=0.265(4). This value is in
units $\mu_c^{-1}$ and is based on a hadronic pion mass of $\mu_c$.
From the two loop values in Table \ref{tab:1} we obtain
\begin{equation}
a^0(\mu_0)-a^2(\mu_0)=a^0(\mu_c)-a^2(\mu_c)-0.0143 \, \rm fm
\label{eq:32}
\end{equation}
If we convert the Colangelo et al. value to fm and assume that
Eq.(\ref{eq:32}) is also true for these results, we then obtain
\begin{equation}
a_{0c}^h(\mu_0)=-\frac{\sqrt{2}}{3}(a^0(\mu_0)-a^2(\mu_0))=-0.170(3) \, \rm fm
\label{eq:33}
\end{equation}

Ref.\cite{18} gives an error for  $a_{0c}^{h}(\mu_c)$ of only 1.8 $\%$.
However, it needs to be remembered that the calculations were made within
the framework of an isospin invariant effective hadronic theory in which
the pions have the mass $\mu_c$ and that the phenomenological data used
have not been corrected in any way for the presence of the electromagnetic
interaction. The same remarks also apply to the determination of the low
energy constants needed to make numerical predictions from ChPT. The
error of  1.8 $\%$ could therefore be an underestimate.

The value  $a_{0c}^{h}(\mu_0)=-0.170(3)$ fm, combined with the result in
Eq.(\ref{eq:26}), leads via Eq.(\ref{eq:11}) to the value $\tau=2.9$ fs,
which agrees with the result of Ref.\cite{14}. We regard this value as an
indication of the lifetime that the DIRAC experiment is measuring.
The important result of Ref.\cite{17} and the present paper is that the
electromagnetic correction in Eq.(\ref{eq:26}) is now determined with
sufficient precision for a measurement of $\tau$ with an accuracy of
10 $\%$ to give a value of the hadronic quantity $a_{0c}^{h}(\mu_0)$ with
an accuracy of not much more than 5 $\%$. We have also emphasised that
true hadronic information needs to be obtained with a pion mass $\mu_0$.
In particular, this mass needs to appear in the formal results of
hadronic ChPT. Also, the phenomenological hadronic data used to fix the
low energy constants and, at a later stage, to refine the hadronic
scattering lengths, need to be corrected for the effect of the
electromagnetic interaction. This is an important long-term programme
and until it is implemented our knowledge of the hadronic scattering
lengths will remain limited. On the other hand, since the electromagnetic
correction given in Eq.(\ref{eq:26}) is now well determined, the
DIRAC experiment will give a clean value of $a_{0c}^{h}(\mu_0)$,
whose error comes almost entirely from the error in the experimental
result.

\begin{ack} 

We thank G. Colangelo for providing us with the phase shifts and scattering
lengths predicted by two-loop ChPT and for helpful discussions. We also
thank J. Gasser and A.Rusetsky for valuable comments on an earlier draft
of our paper. 
We are indebted to the Swiss National Foundation for financial support.

\end{ack}

\end{document}